\documentclass[10pt,conference,compsocconf]{IEEEtran}
\usepackage{times}

\usepackage{caption}
\usepackage{booktabs}
\usepackage[hidelinks]{hyperref}
\usepackage{graphicx}
\usepackage{blkarray,bigstrut}

\ifCLASSINFOpdf
\else
\fi
\begin{document}
\pagenumbering{gobble}

\title{\textbf{\Large Don't Wait to be Breached!\\[-1,5ex]
		Creating Asymmetric Uncertainty of Cloud Applications via Moving Target Defenses}\\[0.2ex]}

\author{\IEEEauthorblockN{~\\[-0.4ex]\large Kennedy A. Torkura\\[0.3ex]\normalsize}
	\IEEEauthorblockA{Hasso Plattner Institute \\University of Potsdam, Germany\\
		Email: {\tt kennedy.torkura@hpi.de}}
	\and
	\IEEEauthorblockN{~\\[-0.4ex]\large Christoph Meinel\\[0.3ex]\normalsize}
	\IEEEauthorblockA{Hasso Plattner Institute \\University of Potsdam, Germany\\
		{\tt christoph.meinel@hpi.de}}
	\and
	\IEEEauthorblockN{~\\[-0.4ex]\large Nane Kratzke\\[0.3ex]\normalsize}
	\IEEEauthorblockA{L\"ubeck University of Applied Sciences\\
		L\"ubeck, Germany\\
		{\tt nane.kratzke@th-luebeck.de}}}

\IEEEspecialpapernotice{(Invited Paper)}

\maketitle

\begin{abstract}
Cloud applications expose – besides service endpoints – also potential or actual vulnerabilities. Therefore, cloud security engineering efforts focus on hardening the fortress walls but seldom assume that attacks may be successful. At least against zero-day exploits, this approach is often toothless.
Other than most security approaches and comparable to biological systems we accept that defensive ``walls" can be breached at several layers. Instead of hardening the ``fortress" walls we propose to make use of an (additional) active and adaptive defense system to attack potential intruders - an immune system that is inspired by the concept of a moving target defense. This ``immune system" works on two layers. On the infrastructure layer, virtual machines are continuously regenerated (cell regeneration) to wipe out even undetected intruders. On the application level, the vertical and horizontal attack surface is continuously modified to circumvent successful replays of formerly scripted attacks.  
Our evaluations with two common cloud-native reference applications in popular cloud service infrastructures (Amazon Web Services, Google Compute Engine, Azure and OpenStack) show that it is technically possible to limit the time of attackers acting undetected down to minutes. Further, more than 98\% of an attack surface can be changed automatically and minimized which makes it hard for intruders to replay formerly successful scripted attacks. So, even if intruders get a foothold in the system, it is hard for them to maintain it.
\end{abstract}

\begin{IEEEkeywords}
zero-day; exploit; moving target defense; microservice; cloud-native; application; security; asymmetric
\end{IEEEkeywords}
\section{Introduction}

\begin{table}[b]
    \caption{\textbf{Some popular open source elastic platforms}}
    \label{tab:platforms} 
    \centering
    \scriptsize
    \begin{tabular}{lll}
        \toprule
        \textbf{Platform} & \textbf{Contributors} & \textbf{URL}\\
        \midrule
        Kubernetes & Cloud Native Found. &  \url{http://kubernetes.io}\\
        Swarm         & Docker & \url{https://docker.io} \\
        Mesos         & Apache & \url{http://mesos.apache.org/} \\
        Nomad     & Hashicorp & \url{https://nomadproject.io/} \\
        \bottomrule
    \end{tabular}
\end{table}

\begin{figure*}
    \centering%
    \includegraphics[width=0.7\textwidth]{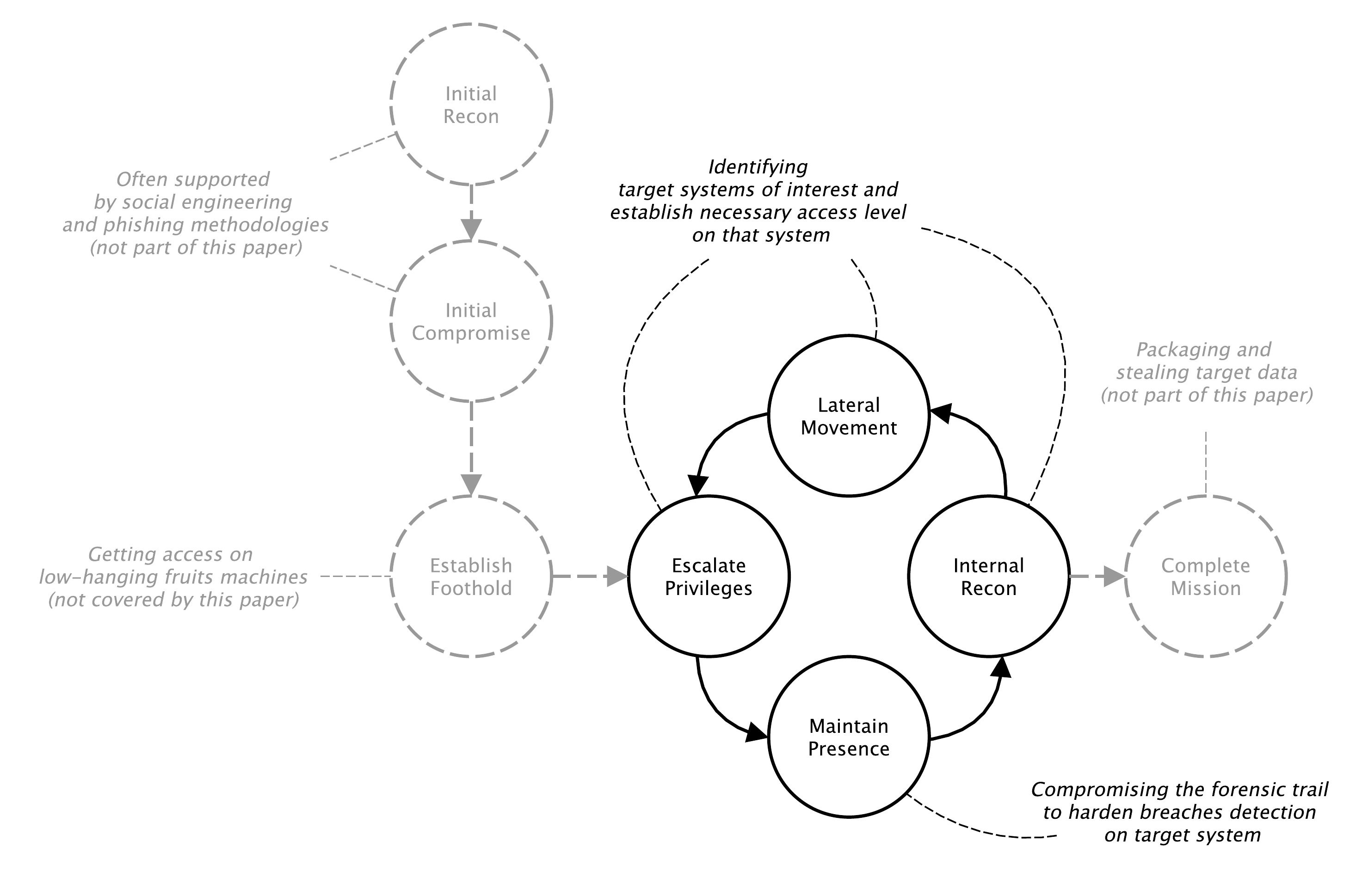}
    
    \caption{\textbf{The cyber attack life cycle model.} \textit{Adapted from the cyber attack lifecycle used by the M-Trends reports, see Table \ref{tab:dwells}}\label{fig:cyber-attack}.}
\end{figure*}

This paper extends ideas presented in \cite{Kra2018b} to improve cloud application security in the context of unknown zero-day exploits and reports on ongoing research in this field.
Cloud computing enables a variety of innovative IT-enabled business and service models, and many research studies and programs focus on responsibly developing systems to ensure the security and privacy of users. But compliance with standards, audits, and checklists, does not automatically equal security \cite{DW2014} and there is a fundamental issue remaining. 
Zero-day vulnerabilities are computer-software vulnerabilities that are unknown to those who would be interested in mitigating the vulnerability (including the entity responsible for operating a cloud application). Until a vulnerability is mitigated, hackers can exploit it to adversely affect computer programs, data, additional computers or a network. For zero-day exploits, the probability that vulnerabilities are patched is zero, so the exploit should always succeed. Therefore, zero-day attacks are a severe threat, and we have to draw a scary conclusion: \textbf{In principle, attackers can establish footholds in our systems whenever they want.}

 This contribution deals with the question how to build ``unfair" cloud systems that permanently jangle attackers nerves. We present the latest results from our ongoing research that applies Moving Target Defense (MTD) principles on cloud runtime environment and cloud application layer.

Recent research \cite{Kra2017a,Kra2018} made successfully use of elastic container platforms (see Table \ref{tab:platforms}) and their ``designed for failure" capabilities to realize transferability of cloud-native applications at runtime. By transferability, the conducted research means that a cloud-native application can be moved from one IaaS provider infrastructure to another without any downtime. These platforms are more and more used as distributed and elastic runtime environments for cloud-native applications \cite{KQ2017a} and can be understood as a kind of cloud infrastructure unifying middleware \cite{KP2016}. It should be possible to make use of the same features to immunize cloud applications simply by moving an application within the same provider infrastructure. To move anything from A to A makes no sense at first glance. However, let us be paranoid and aware that with some probability and at a given time, an attacker will be successful and compromise at least one virtual machine \cite{BD2012}. A transfer from A to A would be an effective countermeasure -- because the intruder immediately loses any hijacked machine that is moved. To understand that, the reader must know that our approach does not effectively move a machine, it regenerates it. To move a machine means to launch a compensating machine unknown to the intruder and to terminate the former (hi-jacked) machine. Whenever an application is moved its virtual machines are regenerated. Moreover, this would effectively eliminate undetected hi-jacked machines. 

However, attackers can run automated attacks against regenerated machines that will incorporate the same set of vulnerabilities. Therefore, this extended paper shows how we further can improve the regenerating security measure by employing MTD at the \textit{application layer} to change the attack surface of the application itself to let even automated and formerly successful attack scripts fail (at least partly). Primarily, this is achieved by diversifying the application in a way that its containerized components are dynamically transformed at \textit{runtime}. The two abstraction layers that compose microservice applications (application layer and the container image layers) are dynamically changed by changing the programming languages of the applications, and consequently, the container images are built to conform to the requirements of the corresponding applications. This combined approach is enforced at runtime to transform the attack surface of cloud-native applications, thereby reducing the possibility of successful attacks. 

The remaining of this paper is \textbf{outlined} as follows: Section \ref{sec:lifecycle-cyber-attacks} presents a cyber-attack lifecycle model to show where our approach intends to break the continuous workflow of security breaches. Section \ref{sec:regenerating-runtime-environments} presents an approach on how MTD can be applied on cloud runtime environment (infrastructure) level to regenerate the \textit{"infrastructure cells"} of a system continuously, leveraging the inherent "designed-for-failure" capabilities of modern container platforms like Kubernetes, Swarm, or Mesos. This continuing regeneration will wipe out even undetected attackers in a system. However, attackers might recognize that they periodically loose foothold in a hi-jacked system and might try to automatize their breaches. To overcome this, Section \ref{sec:mtd-microservices} will present how even the attack surface of an application can be continuously changed and therefore extends our ideas shown in \cite{Kra2018b}. We have to consider that our approach has some limitations. We will discuss these limitations in Section \ref{sec:critical-discussion} and present corresponding related work in Section \ref{sec:related-work}. We conclude our findings in Section \ref{sec:conclusion}.

\section{Cyber Attack Reference Model}
\label{sec:lifecycle-cyber-attacks}

\noindent Figure \ref{fig:cyber-attack} shows the cyber attack life cycle model which is used by the M-Trends reports\footnote{\url{http://bit.ly/2m7UAYb} (visited 9th Nov. 2017)} to report developments in cyber attacks over the years. According to this model, an attacker passes through different stages to complete a cyber attack mission. It starts with initial reconnaissance and compromising of access means. Social engineering methodologies \cite{KHH+2015}  and phishing attacks \cite{GSK2016} very often supports these steps. Intruders aim to establish a foothold near the target.  All these steps are not covered by this paper, because technical solutions are not able to harden the weakest point in security -- the human being.  
The following steps of this model are more important for this paper. According to the life cycle model, the attacker's goal is to escalate privileges to get access to the target system. Because this leaves trails on the system which could reveal a security breach, the attacker is motivated to compromise this forensic trail. According to security reports, attackers make more and more use of counter-forensic measures
to hide their presence and impair investigations. These reports refer to
batch scripts used to clear event
logs and securely delete arbitrary files.
The technique is simple, but the intruders'
knowledge of forensic artifacts
demonstrate increased sophistication, as well as
their intent to persist in the environment. 
With a barely detectable foothold, the internal reconnaissance of the victim's network is carried out to allow the lateral movement to the target system. This process is a complex and lengthy process and may even take weeks. So, infiltrated machines and application components have worth for attackers and tend to be used for as long as possible. Table \ref{tab:dwells} shows how astonishingly many days on average an intruder has access to a victim system. So, basically there is the requirement, that \textbf{(1) an undetected attacker should lose access to compromised nodes of a system as fast as possible.} Furthermore there is the requirement, that it \textbf{(2) must be hard for an attacker to regain foothold in a system by automating successful attacks}. However, how?

Section \ref{sec:regenerating-runtime-environments} will deal with the \textbf{(1) requirement} showing that it is possible to regenerate possibly compromised infrastructure continuously even to get rid of undetected attackers. Section \ref{sec:mtd-microservices} will deal with the \textbf{(2) requirement} and demonstrate that it is possible to change attack surfaces of applications in a way that successful attacks cannot be repeated 1:1.

\begin{table}[t]
	\caption{\textbf{Undetected days on victim systems} \textit{reported by M-Trends. External and internal discovery data is reported since 2015. No data could be found for 2011.}} 
	\label{tab:dwells} 
	\centering
	\scriptsize
	\begin{tabular}{lccr}
		\toprule
		\textbf{Year} & \textbf{External notification} & \textbf{Internal discovery} & \textbf{Median}\\
		\midrule
		2010 & - & - & 416 \\
		2011 & - & - & ? \\
		2012 & - &  - & 243 \\
		2013  & - & - & 229 \\
		2014  & - & - & 205 \\
		2015  & 320 & 56 & 146 \\
		2016  & 107 & 80 & 99 \\
		\bottomrule
	\end{tabular}
\end{table}
\section{Moving Target Defense Mechanisms on the Container Runtime Environment Level}
\label{sec:regenerating-runtime-environments}

Our recent research dealt \cite{KQ2018b} mainly with vendor lock-in and the question how to design cloud-native applications that are transferable between different cloud service providers. One aspect that can be learned from this is that there is no common understanding of what a cloud-native application is. A kind of software that is \textit{``intentionally designed for the cloud"} is an often heard but empty phrase. However, noteworthy similarities exist between various viewpoints on \textit{cloud-native applications} (CNA) \cite{KQ2017a}. A common approach is to define maturity levels in order to categorize different kinds of cloud applications (see Table \ref{tab:camm}).
\cite{FLR+2014} proposed the IDEAL model for CNAs. A CNA should strive for an \textbf{\underline{i}solated state}, is \textbf{\underline{d}istributed}, provides \textbf{\underline{e}lasticity} in a horizontal scaling way, and should be operated on \textbf{\underline{a}utomated deployment machinery}. Finally, its components should be \textbf{\underline{l}oosely coupled}.

\cite{BHJ2015} stress that these properties are addressed by cloud-specific architecture and infrastructure approaches like \textbf{Microservices} \cite{Newman2015}, \textbf{API-based collaboration}, adaption of \textbf{cloud-focused patterns} \cite{FLR+2014}, and \textbf{self-service elastic platforms} that are used to deploy and operate these microservices via self-contained deployment units (containers). Table \ref{tab:platforms} lists some of these platforms that provide additional operational capabilities on top of IaaS infrastructures like automated and on-demand scaling of application instances, application health management, dynamic routing and load balancing as well as aggregation of logs and metrics \cite{KQ2017a}.

\subsection{Regenerating cloud application runtime environments continuously}
\label{sec:transferable-systems}

\begin{figure*}
	\begin{center}
		\includegraphics[width=0.75\textwidth]{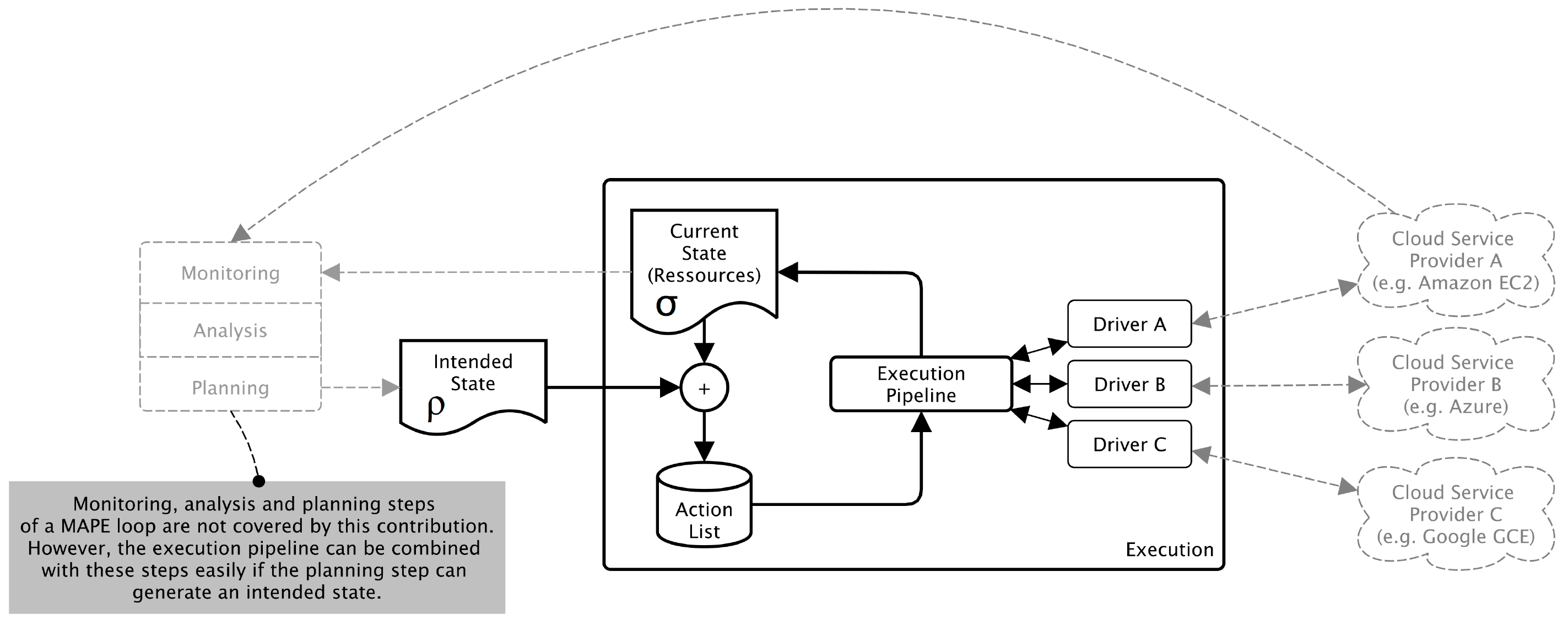}
	\end{center}
	\caption{\textbf{The control theory inspired execution control loop} \textit{compares the intended state $\rho$ of an elastic container platform with the current state $\sigma$ and derives necessary scaling actions. These actions are processed by the execution pipeline explained in Figure \ref{fig:pipeline}. So, platforms can be  operated elastically in a set of synchronized IaaS infrastructures. Explained in details by \cite{Kra2017a}.}}
	\label{fig:executionloop}        
\end{figure*}

\begin{figure*}
	\centering
	\includegraphics[width=0.75\textwidth]{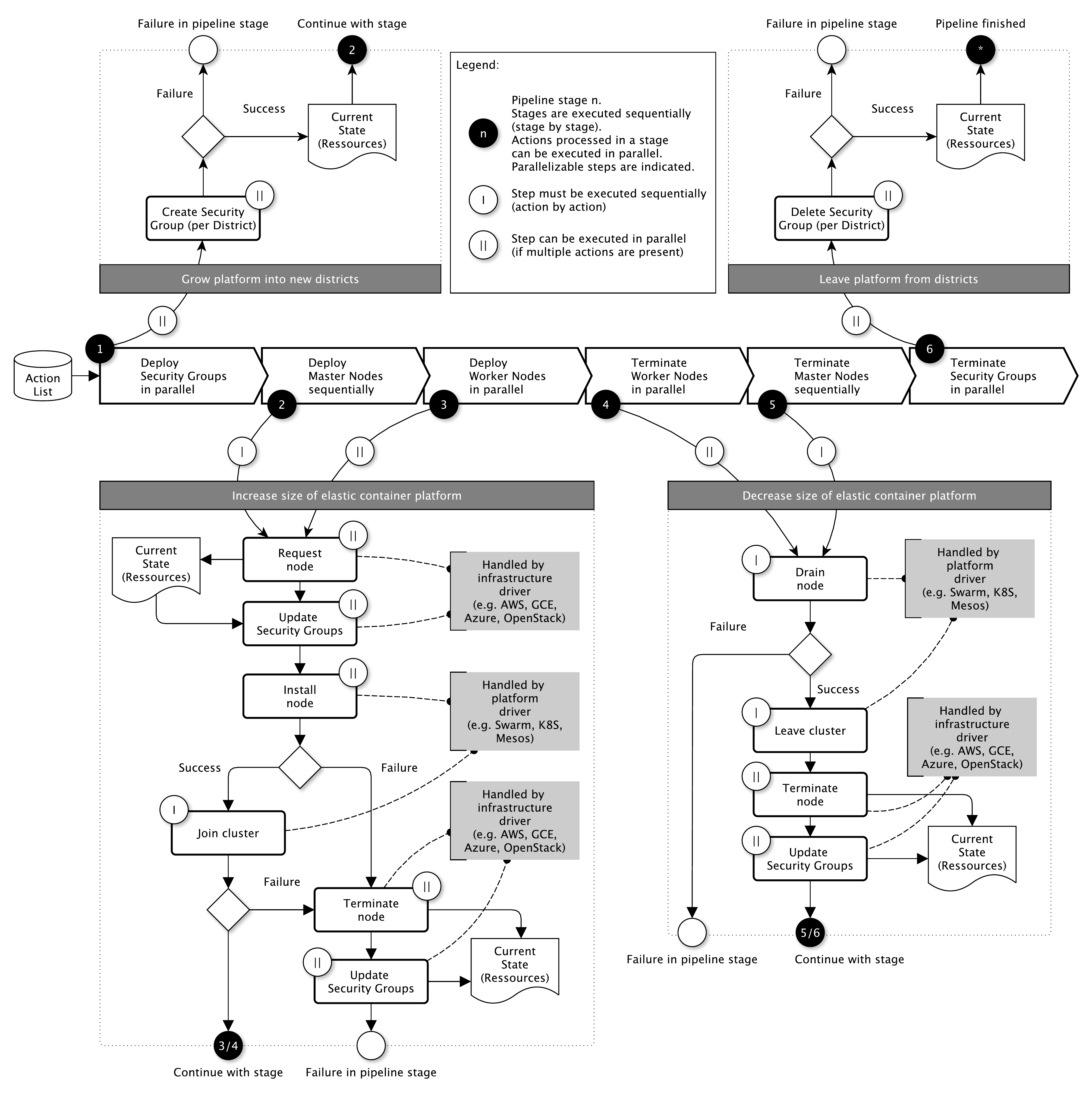}
	\caption{\textbf{The execution pipeline} \textit{processes necessary actions to transfer the current state $\sigma$ into the intended state $\rho$. See \cite{Kra2018} for more details.}}
	\label{fig:pipeline}
\end{figure*}

\begin{table}[b]
	\caption{\textbf{Cloud Application Maturity Model}, adapted from \textit{OPEN DATA CENTER ALLIANCE Best Practices \cite{ODCA2014}}}
	\label{tab:camm} 
	\centering
	\scriptsize
	\begin{tabular}{ccl}
		\toprule
		\textbf{Level} & \textbf{Maturity} & \textbf{Criteria}\\
		\midrule
		3 & Cloud     & - Transferable across infrastructure providers at \\
		& native    &   \enspace runtime and without interruption of service.\\
		&           & - Automatically scale out/in based on stimuli.\\
		\midrule
		2 & Cloud     & - State is isolated in a minimum of services. \\
		& resilient & - Unaffected by dependent service failures. \\
		&           & - Infrastructure agnostic.\\
		\midrule
		1 & Cloud     & - Composed of loosely coupled services. \\
		& friendly  & - Services are discoverable by name. \\
		&           & - Components are designed to cloud patterns. \\
		&           & - Compute and storage are separated. \\
		\midrule
		0 & Cloud     & - Operated on virtualized infrastructure. \\
		& ready     & - Instantiateable from image or script. \\
		\bottomrule
	\end{tabular}
\end{table}

If the reader understands and accepts the commonality that cloud-native applications are operated (more and more often) on elastic -- often container-based -- platforms, it is an obvious idea to delegate the responsibility to immunize cloud applications to these platforms. Recent research showed that the operation of these elastic container platforms and the design of applications running on top of them should be handled as two different engineering problems. This point of view often solves several issues in modern cloud-native application engineering \cite{Kra2018}. Also, that is not just true for the transferability problem but might be an option to tackle zero-day exploits. These kinds of platforms could be an essential part of the immune system of modern cloud-native applications.

Furthermore, \textbf{self-service elastic platforms} are really ``bulletproofed" \cite{Stine2015}. \textit{Apache Mesos} \cite{mesos} has been successfully operated for years by companies like Twitter or Netflix to consolidate hundreds of thousands of compute nodes. Elastic container platforms are \textbf{designed for failure} and provide self-healing capabilities via auto-placement, auto-restart, auto-replication, and auto-scaling features. They will identify lost containers (for whatever reasons, e.g., process failure or node unavailability) and will restart containers and place them on remaining nodes. These features are necessary to operate large-scale distributed systems resiliently. However, the same features can be used intentionally to \textbf{purge ``compromised nodes"}.

\cite{Kra2017a} demonstrated a software prototype that provides the control process shown in Figure \ref{fig:executionloop} and Figure \ref{fig:pipeline}. This process relies on an \textit{intended state} $\rho$ and a \textit{current state} $\sigma$ of a container cluster. If the intended state differs from the current state ($\rho\not = \sigma$), necessary adaption actions are deduced (creation and attachment/detachment of nodes, creation and termination of security groups) and processed by an execution pipeline fully automatically (see Figure \ref{fig:pipeline}) to reach the \textit{intended state} $\rho$. With this kind of control process, a cluster can be simply resized by changing the intended amount of nodes in the cluster. If the cluster is shrinking and nodes have to be terminated, affected containers of running applications will be rescheduled to other available nodes.

The downside of this approach is, that this will only work for Level 2 (cloud resilient) or Level 3 (cloud-native) applications (see Table \ref{tab:camm}) which by design, can tolerate dependent service failures (due to node failures and container rescheduling). However, for that kind of Level 2 or Level 3 application, we can use the same control process to regenerate nodes of the container cluster. The reader shall consider a cluster with $\sigma=N$ nodes. If we want to regenerate one node, we change the intended state to $\rho=N+1$ nodes which will add one new node to the cluster ($\sigma'=N+1$). Moreover, in a second step, we will decrease the predetermined size of the cluster to $\rho'=N$ again, which affects that one node of the cluster is terminated ($\sigma''=N$). So, a node is regenerated simply by adding one node and deleting one node. We could even regenerate the complete cluster by changing the cluster size in the following way: $\sigma = N \mapsto  \sigma' = 2N \mapsto \sigma'' = N$. However, this would consume much more resources because the cluster would double its size for a limited amount of time. A more resource efficient way would be to regenerate the cluster in $N$ steps: $\sigma = N \mapsto  \sigma' = N+1 \mapsto \sigma'' = N \mapsto ... \mapsto \sigma^{2N-1}=N+1 \mapsto \sigma^{2N}=N$.    The reader is referred to \cite{Kra2018} for more details, especially if the reader is interested in the multi-cloud capabilities, that are not covered by this paper due to page limitations.

Whenever such regeneration is triggered, all -- even undetected -- hijacked machines would be terminated and replaced by other machines, but the applications would be unaffected. For an attacker, this means losing their foothold in the system entirely. Imagine this would be done once a day or even more frequently?

\subsection{Evaluation}
\label{subsec:runtime-evaluation}

\begin{table}[b]
	\caption{\textbf{Used machine types and regions for evaluation}}
	\label{tab:machine-types} 
	\centering
	\scriptsize
	\begin{tabular}{llll}
		\toprule
		\textbf{Provider} & \textbf{Region} &\textbf{Master type} & \textbf{Worker type}\\
		\midrule
		AWS & eu-west-1 & m4.xlarge & m4.large \\
		GCE & europe-west1 & n1-standard-4 & n1-standard-2 \\
		Azure & europewest & Standard\_A3 & Standard\_A2 \\
		OS & \textit{own datacenter} & m1.large & m1.medium \\
		\bottomrule
	\end{tabular}
\end{table}

\begin{table}[b]
	\caption{\textbf{Durations to regenerate a node (median values)}}
	\label{tab:regeneration-times} 
	\centering
	\scriptsize
	\begin{tabular}{lrrrrr}
		\toprule
		\textbf{Provider} & \textbf{Creation} &\textbf{Secgroup} & \textbf{Joining} & \textbf{Term.} & \textbf{\underline{Total}}\\
		\midrule
		AWS & 70 s & 1 s & 7 s & 2 s & \textbf{81 s} \\
		GCE & 100 s & 8 s & 9 s & 50 s & \textbf{175 s} \\
		Azure & 380 s & 17 s & 7 s & 180 s & \textbf{600 s} \\
		OS  & 110 s & 2 s & 7 s & 5 s & \textbf{126 s} \\
		\bottomrule
	\end{tabular}
\end{table}

\noindent The execution pipeline presented in Figure \ref{fig:pipeline} was evaluated by operating and transferring two elastic platforms (\textit{Swarm Mode of Docker 17.06} and \textit{Kubernetes 1.7}). The platforms operated a reference ``sock-shop" application being one of the most complete reference applications for microservices architecture research \cite{AMP+2017}. Table \ref{tab:machine-types} lists  the machine types that show a high similarity across different providers \cite{KQ2015}.

\begin{figure}[t]
	\centering
	\includegraphics[width=\columnwidth]{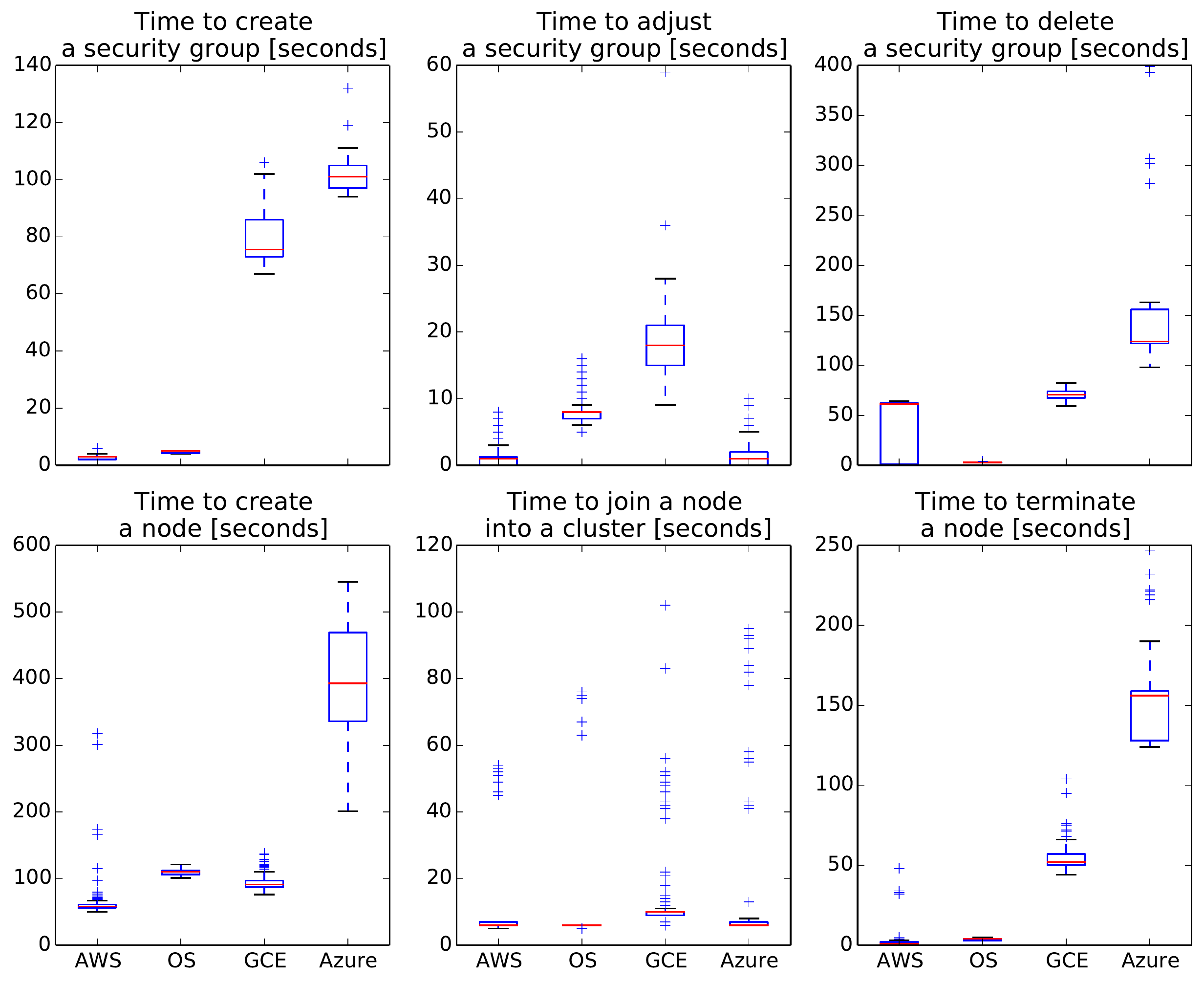}
	\caption{\textbf{Infrastructure specific runtimes of IaaS operations} \textit{see \cite{Kra2018}.}}
	\label{myfigone}
\end{figure}

The evaluation of \cite{Kra2018} demonstrated that most time is spent on the IaaS level (creation and termination of nodes and security groups) and not on the elastic platform level (joining, draining nodes). The measured differences on infrastructures provided by different providers are shown in Figure \ref{myfigone}. For the current use case, the reader can ignore the times to create and delete a security group (because that is a one time action). However, there will be many node creations and terminations. According to our execution pipeline shown in Figure \ref{fig:pipeline}, a node creation ($\sigma = N \mapsto \sigma'=N+1$) involves the durations to \textbf{create a node} (request of the virtual machine including all installation and configuration steps), to \textbf{adjust security groups} the cluster is operated in and to \textbf{join the new node} into the cluster. The shutdown of a node ($\sigma=N \mapsto \sigma'=N-1$) involves the \textbf{termination of the node} (this includes the platform draining and deregistering of the node and the request to terminate the virtual machine) and the necessary \textbf{adjustment of the security group}. So, for a complete regeneration of a node ($\sigma=N \mapsto \sigma'=N+1 \mapsto \sigma''=N$) we have to add these runtimes. Table \ref{tab:regeneration-times} lists these values per infrastructure.

Even on the ``slowest" infrastructure, a node can be regenerated in about 10 minutes. In other words, one can regenerate six nodes every hour or up to 144 nodes a day or a cluster of 432 nodes every 72h (which is the reporting time requested by the EU General Data Protection Regulation). If the reader compares a 72h regeneration time of a more than 400 node cluster (most systems are not so large) with the median value of 99 days that attackers were present on a victim system in 2016 (see Table \ref{tab:dwells}) the benefit of the proposed approach should become apparent.
\section{Moving Target Defense Mechanisms on the Microservice Architecture Level}
\label{sec:mtd-microservices}


\label{subsec:design-and-system-model} MTD techniques introduce methods for improving the security of protected assets by applying \textit{security-by-diversity} tactics and \textit{security diversification} concepts. While most MTD techniques do not have formal requirements for diversifying, i.e. when, how and why to diversify, we employ a \textit{cyber risk-based} technique as the primary diversification decision making factor on the application level. Our motivation for this is to overcome the high number of vulnerability infection among container images as shown by several recent researchers\cite{gummaraju2015over,shu2017study}. Therefore, our MTD techniques are designed to improve this \textit{state of insecurity} by reducing the \textit{window of vulnerability exposure} via diversification and commensurate attack surface randomization.

\subsection{Cyber Risk Analysis for Microservice Diversification}\label{subsec:secriskmetric} Larsen et al. \cite{larsen2015automated}  assert that a common challenge when employing diversification strategies is deciding on \textit{when}, \textit{how} and \textit{where} to diversify. We present a cyber risk procedure to support decision making or satisfy the above afore-mentioned requirements. We leverage security metrics to design a cyber risk-based mechanism, and security metrics are useful tools for risk assessment. These metrics are computed by deriving security risks per microservice and after that employing \textit{vulnerability prioritization} such that diversification is a function of microservice risk assessment, i.e. microservices are diversified in order risk severity. We introduce the notion of \textbf{Diversification Index - $D_i$} as an expression of the depth of diversification to be implemented.  $D_i$ defines if microservices are to be globally or selectively diversified. Diversifying 2 out of 4 microservices can be expressed as \textit{2:4}.  $D_i$ is formally defined as:

\begin{equation}
D_i = \frac{m_d}{m}
\end{equation}

where,

$m_d$ =  number of microservices to be diversified,

$m$ = total number of microservices in the application. For this, we adopt two approaches:

\subsubsection{Risk Analysis Using CVSS}\label{subsubsec:cvss} The Common Vulnerability Scoring System {CVSS} \cite{mell2006common} is a widely adopted vulnerability metrics standard. It provides vulnerability \textit{base scores} which express the severity of damage the referred vulnerability might impact upon a system if exploited. In order to derive the microservice security state (Security Risk - $SR$), base scores of all the vulnerabilities detected can be summed and averaged as expressed below:

\begin{equation}
\label{eqnTwo}
SR = \frac{1}{N}\sum_{i=1}^{N} V_i
\end{equation}
where $SR$ is the Security Risk, $V_i$ is the CVSS base score of vulnerability $i$, and $N$ is the total number of vulnerabilities detected in microservice $m$. However, averaging vulnerabilities to obtain a single metric to signify a system's security state is not optimal. Derived values are not sufficiently representative of other factors such as the public availability of exploits. Therefore, we employ another scoring technique called \textit{shrinkage estimator}, an approach which has been popularly used for online rating systems, e.g. IMDB. The shrinkage estimator considers the average rating and the number of votes. Hence, it provides a more precise value for SR, than mere averaging (Equation \ref{eqnTwo}). Therefore, leveraging the shrinkage estimator, we can derive a more precise $SR$ as follows:

\begin{equation}
\label{eqnThree}
SR = \frac{v}{v+a}  R + \frac{a}{v+a} C
\end{equation}

where, 

$v$ = the total number of vulnerabilities detected in a microservices,

$a$ = minimum number of vulnerabilities to be detected in a microservice assessment before it added in the risk analysis,

$C$ = the mean severity score of vulnerabilities detected in a microservice

$R$ = the average severity score of all vulnerabilities infecting a microservice-based application


The Pearson's correlation coefficient is derived to determine the dependence relationship between the microservices.  

\subsubsection{Risk Analysis Using OWASP Risk Rating Methodology}\label{subsubsec:owasp}
The risk assessment method described in the previous subsection is limited to vulnerabilities contained in the Common Vulnerability Enumeration (CVE) dictionary. CVE is a public dictionary for publishing known vulnerabilities. These vulnerabilities are analyzed and assigned vulnerability security metrics using the CVSS. However, the CVE contains only a handful of web application vulnerabilities. Thus we need to derive another risk assessment methodology for application layer vulnerabilities. This additional step is necessary since microservices are essentially web/REST-based applications. We opt for the OWASP Risk Rating Methodology (ORRM), which is specifically designed for web applications \cite{OWASPRiskRatingMethodology}. This methodology is based on two core risk components: \textit{Likelihood} and \textit{Impact} formally expressed as:
\begin{equation}
Risk = Likelihood * Impact
\end{equation}
In order to derive these metrics,  risk assessors are required to consider the threat vector, attacks to be used and the impacts of successful attacks.
\begin{figure}  
	\center
	\includegraphics[width=0.50\textwidth]{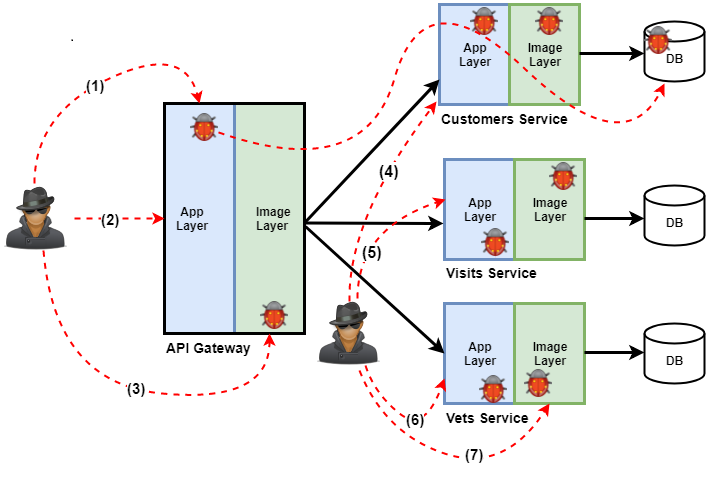}
	\caption{Typical Microservice Attack Surfaces illustrated with the PetClinic Application \cite{spring_petclinic}}\label{fig:microattksurfaces}
\end{figure}

\subsection{Dissecting Microservice Attack Surfaces}\label{subsec:multicorrelation}
An important aspect of our \textit{security-by-diversity} tactics is to manipulate microservice attack surfaces against possible attackers through random architectural transformations. Therefore, the attack surfaces are altered by randomizing the entry and exit points, which are commonly used for identifying attack surfaces \cite{younis2016assessing,manadhata2009report}. A detailed understanding of these attack surfaces is imperative. Therefore, we categorize microservice attack surface into: \textit{horizontal} and \textit{vertical} attack surfaces and thereafter employ \textit{vulnerability correlation} to identify vulnerability similarities.

\subsubsection{Horizontal Vulnerability Correlation}\label{subsubsec:horicorrelation} The objective of correlating vulnerabilities horizontally is to analyze the relationship of vulnerabilities along the horizontal attack surface, i.e. the parts of the applications users directly interact with. Figure \ref{fig:microattksurfaces} illustrates the multi-layered attack surface of the PetClinic application \cite{spring_petclinic}. The application layer horizontal attack surface consists of the interactions and exit/entry points from the API gateway to the Vets, Visits and Customer services application layers. Requests and responses are transversed along this layer, providing attack opportunities for attackers. The vulnerability correlation process is similar to \textit{security event correlation} techniques \cite{ficco2013security}, though rather than clustering similar attributes e.g., malicious IP addresses, we focus on Common Weakness Enumeration (CWE) Ids. The CWE is a standardized classification system for application weaknesses\footnote{https://cwe.mitre.org/index.html}. For example, CWE 89  categorizes all vulnerabilities related to \textit{Improper Neutralization of Special Elements used in an SQL Command (SQL Injection)}\footnote{https://cwe.mitre.org/data/definitions/89.html} and can be mapped to several CVEs e.g. \textit{CVE-2016-6652}\footnote{https://nvd.nist.gov/vuln/detail/CVE-2016-6652}, a SQL injection vulnerability in Spring Data JPA. If this vulnerability exists in all PetClinic's microservices, an attacker could easily conduct a correlated attack (\textit{Attack Paths 2, 4, 5, and 6} of Figure \ref{fig:microattksurfaces}) resulting to correlated failures and eventual application failure since each microservice works ultimately to the successful functioning of the PetClinic application.

\subsubsection{Vertical Vulnerability Correlation}\label{subsubsec:vercorrelation}
The vertical correlation technique is similar to the horizontal correlation. However, the interactions across application-image layers are analyzed. This analysis, therefore, employs security-by-design tactics across the vertical attack surface. Attack Path 1 illustrates the exploitation of vulnerability across the vertical attack surface, and the attacker initiated an attack against the API Gateway of the PetClinic application, from the application layer to the image layer. From there, another attack is launched to the Customers service application layer, across the image layer and finally, the database is compromised. The same attack can be repeated against the other microservices if affected by the vulnerabilities. Hence we need to express such casual relationships in vulnerability correlation matrices.

\begin{figure}[!ht]
	\[
	\begin{blockarray}{ccccc}
	& V_1 & V_2 & \dots & V_n \\
	\begin{block}{c[cccc]}
	M_1 & 1&1&\cdots &\dots\bigstrut[t] \\
	M_2 & 1&0&\cdots &\dots \\
	& \vdots & \vdots & \ddots & \vdots\\
	M_n & 1&1&\cdots &\dots\bigstrut[b]\\
	\end{block}
	\end{blockarray}\vspace*{-1.25\baselineskip}
	\]
	\caption{Microservice Vulnerability Correlation Matrix}\label{fig:vulnmatrix}
\end{figure}

\label{subsec:microcormatrix}
Correlated vulnerabilities can be represented with correlation matrices, more specifically referred to as \textit{microservices vulnerability correlation matrix}. Therefore, we are influenced by \cite{chen2011correlated} to define the microservices vulnerability correlation matrix as \textit{a mapping of vulnerabilities to microservice instances in a microservice-based application}. The  \textit{microservices vulnerability correlation matrix} presents a view of vulnerabilities that concurrently affect multiple microservices. An example of the microservice correlation matrix is Figure \ref{fig:vulnmatrix}, where the microservices $M_1$ and $M_2$ will have a correlated failure under an attack that exploits vulnerability $V_1$ since they share the same vulnerability. However, an attack that exploits $V_2$ can only affect $M_1$, while $M_2$ remains unaffected.

\subsection{Evaluation}
\label{subsec:microservice-evaluation}
The PetClinic application was used for our evaluation. PetClinic is part of the Spring Cloud demo applications and an established cloud-native reference application used for demonstration purposes in plenty of industrial and academic microservice-related use cases \cite{AMP+2017}. It is, therefore, an excellent reference. However, we were forced to modify the original PetClinic by adding OpenAPI support. Two experiments have been conducted: (1) Security risk comparison to verify the efficiency of our security-by-diversity tactics (2) Attack surface analysis to evaluate the improvement in the horizontal and vertical attack surfaces.

\begin{table}
	\centering
	\caption{Vulnerabilities Detected in PetClinic App-Layer}
	\label{tab:petclinicvulns}
	\resizebox{\columnwidth}{!}{%
		\begin{tabular}{|c|c|c|c|c|}
			\hline
			\textbf{CWE-ID} & \textbf{API-GATEWAY} & \textbf{CUSTOMERS-SERVICE} & \textbf{VETS-SERVICE} & \textbf{VISITS-SERVICE} \\
			\hline
			CWE-16 & 31 & 4 & 2 & 2 \\
			CWE-524 & 48 & 17 & 6 & 11 \\
			CWE-79 & 0 & 3 & 0 & 1 \\
			CWE-425 & 0 & 0 & 20 & 0 \\
			CWE-200 & 14 & 6 & 0 & 0 \\
			CWE-22 & 0 & 1 & 0 & 0 \\
			CWE-933 & 1 & 0 & 0 & 0 \\
			\hline
			\textbf{TOTAL} & 94 & 31 & 28 & 14\\
			\hline
		\end{tabular}
	}
\end{table}

\begin{figure}  
	\center
	\includegraphics[width=0.40\textwidth]{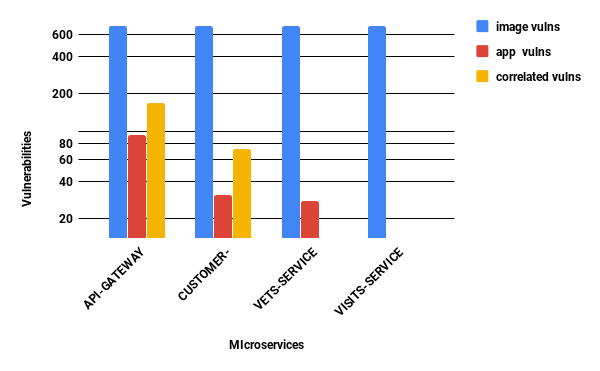}
	\caption{Vulnerability scanning results of the Homogeneous PetClinic application }\label{fig:homogenious_petclininc}
\end{figure}

\label{subsec:riskanalysiseaval}

\begin{figure}  
	\center
	\includegraphics[width=0.40\textwidth]{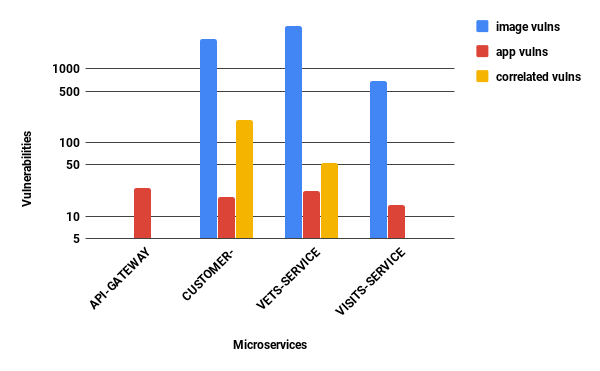}
	\caption{Vulnerabilities detected in the Diversified PetClinic Application }\label{fig:polyglot_petclinic}
\end{figure}

In order to perform Security Risk analysis, we leveraged the Cloud Aware Vulnerability Assessment System (CAVAS) \cite{torkura2018cavas}. The vulnerability scanners integrated into CAVAS (Anchore and OWASP ZAP), are used for launching vulnerability scans against PetClinic images and microservice instances respectively. The detected vulnerabilities were persisted in the Security Reports and CMDB. First, the diversification index is derived by computing risks per PetClinic microservices to obtain the \textit{Security Risk} - $SR$. Hence, we inspect the results for the image vulnerability scan and notice that the vulnerabilities are too similar (Figure \ref{fig:homogenious_petclininc}). Therefore, $SR$ will be too similar for meaningful vulnerability prioritization. Since the prioritization step is imperative for ordering microservices in order of risk severity, we compute $SR$ using the ORRM (Section  \ref{subsubsec:owasp}).  The application layer scan results are retrieved from the database and analyzed. Scores are assigned to the detected vulnerabilities based on the risk scores for OWASP Top-10 2017 web vulnerabilities \cite{owasp10application}. This is a reasonable approach given OWASP uses ORRM for deriving the Top-10 web application vulnerability scores. Also, this affords objective assignment of scores\footnote{https://www.owasp.org/index.php/Top\_10-2017\_Details\_About\_Risk\_Factors}, which are publicly verifiable.  Table \ref{tab:petclinicvulns} is the distribution of detected vulnerabilities, while a subset of the mapping between CWE-Ids and OWASP Top-10 is on Table \ref{tab:riskscoresbycwe}.  From Table \ref{tab:riskscoresbycwe}, it is obvious that the API-Gateway has the most severe risks followed by the Customer, Vets, and Visits microservices. Therefore, we apply diversification based on this result using a \textit{diversification index of 3:4}, i.e. three out of four microservices. The diversified PetClinic is \textit{retested} and the results are shown in Figure \ref{fig:polyglot_petclinic}. We observe that the diversified PetClinic application layer vulnerabilities are reduced with about 53.3 \%. However, the image vulnerabilities increased especially for the Customer and Vets service which are transformed to NodeJS and Ruby respectively. Importantly, the microservices are no longer homogeneous, and the possibilities for correlated attacks have been eliminated. Also, the vulnerabilities in the API Gateway's image are drastically reduced from 696 to 6, while the application layer vulnerabilities reduced from 94 to 24. The reduction is due to reduced code base size, a distinct characteristic of Python programming model. The API Gateway is the most important microservice since it presents the most vulnerable and sensitive attack surface of the application, therefore consider the security of PetClinic improved, our results mean that out of 94 opportunities for attacking the API Gateway, only 24 were left.

\begin{table}
	\centering
	\caption{Risk Scores By CWE }
	\label{tab:riskscoresbycwe}
	\begin{tabular}{| l | c | r |}
		\hline
		\textbf{CWE-ID}  & \textbf{OWASP T10 Risk Category} & \textbf{Risk Score} \\
		\hline
		CWE-16  & A6 - Security Misconfiguration & 6.0        \\
		CWE-524 & Not Listed                     & 3.0        \\
		CWE-79  & A6 - Security Misconfiguration & 6.0        \\
		CWE-425 & Not Listed                     & 3.0        \\
		CWE-200 & A3 - Sensitive Data Exposure   & 7.0        \\
		CWE-22  & A5 - Broken Access Control     & 6.0        \\
		CWE-933 & Not Listed                     & 3.0       \\
		\hline
	\end{tabular}
\end{table}

\begin{figure}  
	\center
	\includegraphics[width=0.40\textwidth]{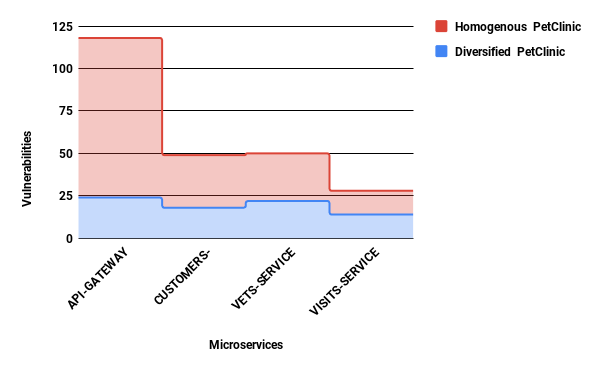}
	\caption{Horizontal Attack Surface Analysis}\label{fig:horattksurface}
\end{figure}

\subsection{Attack Surface Analysis}\label{subsec:attacksurface}

Here we analyze the attack surfaces of the homogeneous and diversified PetClinic versions. We consider direct and indirect attack surfaces, i.e. vulnerabilities that directly/ indirectly lead to attacks respectively. From the vulnerability scan reports, each detected vulnerability is counted as an attack surface \textit{unit} (\textit{attack opportunities concept} \cite{howard2005measuring,owasp_attk_surfaces}). Figure \ref{fig:horattksurface} compares the horizontal app layer attack surface for both PetClinic apps. Notice a reduced attack surface in the diversified version, showing better security. Essentially, the attackability of PetClinic has been reduced. However, the results for the vertical attack surface are different. This attack surface portrays attacks transversing the app-image layer (Figure \ref{fig:microattksurfaces}). While there are fewer correlated vulnerabilities in the diversified API-Gateway, correlated vulnerabilities in the Customers and Vets Services have increased. This increment is due to the corresponding increase of image vulnerabilities. However, the attackability due to homogeneity is reduced. 

We want to emphasize that intruders would observe this approach as permanently changing attack surfaces increasing dramatically the effort to breach the system.

\section{Critical Discussion}
\label{sec:critical-discussion}

\noindent The idea presented in Section \ref{sec:regenerating-runtime-environments} of an immune system like approach to remove undetected intruders in virtual machines seems to a lot of experts intriguing. Nevertheless, according to the state of the art, this is currently not done. There might be reasons for that and open questions the reader should consider.

It is often remarked that the proposal can be compared with the approach to restart periodically virtual machines that have memory leak issues and has apparently nothing to do with security concerns, and could be applied to traditional (non-cloud) systems as well. So, the approach may have even a broader focus than presented (which is not a bad thing).

Another question is how to detect ``infected" nodes? The presented approach selects nodes simply at random and will hit every node at some time. The same could be done using a round-robin approach, but a round-robin strategy would be better predictable for an attacker. However, both strategies will create a lot of additional regenerations, and that leaves room for improvements. It seems obvious to search for solutions like presented by \cite{FLW+2009,WSF+2017} to provide some  ``intelligence" for the identification of  ``suspicious" nodes. Such a kind of intelligence would limit regenerations to likely ``infected" nodes. In all cases, it is essential for anomaly detection approaches to secure the forensic trail \cite{DW2016b,DW2016a}.

Furthermore, to regenerate nodes periodically or even randomly is likely nontrivial in practice and depends on the state management requirements for the affected nodes. Therefore, this paper proposes the approach only as a promising solution for Level 2 or 3 cloud applications (see Table \ref{tab:camm}) that are operated on elastic container platforms. These kinds of applications have desirable state management characteristics. However, this is a limitation to applications following the microservice architecture approach.

One could be further concerned about exploits that are adaptable to bio-inspired systems. Stealthy resident worms dating back to the old PC era would be an example. This concern might be especially valid for the often encountered case of not entirely stateless services when data-as-code dependencies or code-injection vulnerabilities exist. Furthermore, attackers could shift their focus to the platform itself in order to disable the regeneration mechanism as a first step. On the other hand, this could be easily detected -- but there could exist more sophisticated attacks.

These ``immunization" results on the infrastructure level (see Section \ref{sec:regenerating-runtime-environments}) are impressive but should be combined with secure coding practices in development pipelines, i.e. employing with continuous security assessments. We presented how to automate security in CNA development environments \cite{torkura2018cavas}. In these cases, detected web vulnerabilities, e.g. \textit{X-Content-Type-Options Header Missing}, can be resolved by appending appropriate \textit{headers}, as described and advised in CAVAS reports. Furthermore, image vulnerabilities can be reduced by using more secure container images. For example, \textit{Alpine Linux} images can replace Ubuntu images as base images due to smaller footprint which equals smaller attack surfaces \cite{gantikow2016providing}. 

Our MTD approach presented in Section \ref{sec:mtd-microservices} leverages automatic code generation techniques on the application level via Swagger CodeGen library. We discovered that over 150 companies/projects use Swagger CodeGen in production\footnote{https://github.com/swagger-api/swagger-codegen}, hence the library is mature and capable of transforming large microservice applications. Nevertheless, in this work a basic application has been used to introduce the concepts, more complex applications will be tested in the future. However, our approach also has some limitations. Our techniques can be applied only to OpenAPI compatible microservices. Also, Swagger Codegen currently supports about 30 programming languages/frameworks and this might be a limitation in terms for possible combinations (entropy), although more languages can be added via customizations. There might be a need for manual efforts to check if the transformation output is functionally compatible especially for complex applications. 

An event-based technique might interestingly enhance our MTD technique by detecting attacks and triggering commensurate diversification. Conventionally, Web Application Firewalls (WAF) are deployed in front of web applications to detect and stop malicious traffic (which might also indicate an ongoing attack). Hence WAF can be deployed at the API Gateway and configured with \textit{attack thresholds}. Once a threshold is breached, the WAF would trigger the diversification of the entire microservice application or the endangered microservice. A scheduled diversification routine might support this methodology. These techniques can comfortably be applied across cloud platforms using orchestration technologies, e.g. Kubernetes.
\section{Related Work}
\label{sec:related-work}

To the best of the authors' knowledge, there are currently no approaches making intentional use of virtual machine regeneration for security purposes neither on the infrastructure nor on the application level. However, the proposed approach is derived from multi-cloud scenarios and their increased requirements on security. Moreover, several promising approaches are dealing with multi-cloud scenarios. So, all of them could show equal opportunities. However, often, these approaches come along with much inherent complexity. A container-based approach seems to handle this kind of complexity better. There are some good survey papers on this \cite{Barker2015,Petcu2014,Toosi2014,Grozev2014}.

MTD via software diversity was first introduced by Forest et al. \cite{forrest1997building}, since then the concept has been applied at different abstraction levels. Baudry et al. \cite{baudry2014tailored} introduced \textit{sosiefication}, a diversification method which transforms software programs by generating corresponding replicas through statement deletion, addition or replacement operators. These variants still exhibit the same functionality but are computationally diverse.  Williams et. al  \cite{williams2009security} presented \textit{Genesis}, a VM-based dynamic diversification system. Genesis employed the \textit{Strata} VM to distribute software components such that every version became unique, hence difficult to attack. A detailed comparison of automated diversification techniques was presented in \cite{larsen2015automated}. The authors have not found a prior work that applied MTD concepts to microservices.

\section{Conclusion}
\label{sec:conclusion}

There is still no such thing as an impenetrable system. Once attackers successfully breach a system, there is little to prevent them from doing arbitrary harm – but we can reduce the available time for the intruder to do this. Moreover, we can make it harder to replay a successful attack. The presented approach evolved mainly from transferability research questions for cloud-native applications. Therefore, it is limited to microservice-based application architectures but provides some unusual characteristics for thinking about security in general. 

Basically we proposed an ``immune system" inspired approach to tackle zero-day exploits. The founding cells are continuously regenerated. The primary intent is to reduce the time for an attacker acting undetected massively. Therefore, this paper proposed to regenerate virtual machines (the cells of an IT-system) with a much higher frequency than usual to purge even undetected intruders. Evaluations on infrastructures provided by AWS, GCE, Azure, and OpenStack showed that a virtual machine could be regenerated between two minutes (AWS) and 10 minutes (Azure). The reader should compare these times with recent cybersecurity reports. In 2016 an attacker was undetected on a victim system for about 100 days. The presented approach means for intruders that their undetected time on victim systems is not measured in months or days any-more, it would be measured in minutes. 

However, regenerated virtual machines will incorporate the same set of application vulnerabilities. So, a reasonable approach for intruders would be to script their attacks and rerun it merely. Although they might lose their foothold within minutes in a system, they can regain it automatically within seconds. Therefore, we propose to alter the attack surface of applications by randomizing the entry and exit points, which are commonly used for identifying attack surfaces \cite{younis2016assessing,manadhata2009report}. Based on horizontal and vertical microservice attack surfaces we demonstrated how to employ a \textit{vulnerability correlation} to identify vulnerability similarities on the application layer and how to adapt the attack surface accordingly. This attack surface modification would let even automated and formerly successful attack scripts fail (at least partly). We propose and demonstrate the feasibility to diversify the application via dynamic transformations of its containerized components at \textit{runtime}. In our presented use cases, we could show, that it is possible to change the attack surface of a reference application incorporating over 600 container image vulnerabilities and approximately 80 application vulnerabilities to a surface with no image vulnerabilities and only 24 application vulnerabilities anymore. That is a reduction of almost 98\%. What is more, the surface of the application can be changed continuously resulting that scripted attacks fail with each surface change. That is a nightmare from an intruders point of view.

The critical discussion in Section \ref{sec:critical-discussion} showed that there is a need for additional evaluation and room for more in-depth research on both levels: continuously infrastructure regeneration and application surface modifying. However, several reviewers remarked independently that the basic idea is so ``intriguing", that it should be considered more consequently.

\section*{Acknowledgment}
This research is partly funded by the Cloud TRANSIT project (13FH021PX4, German Federal Ministry of  Education and Research). The authors would like to thank Bob Duncan from the University of Aberdeen for his inspiring thoughts on cloud security challenges.

\bibliographystyle{IEEEtran}
\bibliography{ref}

\end{document}